\documentclass[superscriptaddress,showpacs,twocolumn,pra]{revtex4}
\usepackage{xspace,amsmath,amsfonts,amsthm,amssymb,graphicx,color}

\begin{document}

\title{A bound on the mutual information, and properties of entropy reduction, for quantum channels with inefficient measurements}

\author{Kurt Jacobs}
\affiliation{Centre for Quantum Computer Technology, Centre for Quantum Dynamics, School of Science, Griffith University, Nathan 4111, Brisbane, Australia}

\begin{abstract}

The Holevo bound is a bound on the mutual information for a given quantum encoding. 
In 1996 Schumacher, Westmoreland and Wootters [Schumacher, Westmoreland and 
Wootters, Phys. Rev. Lett. {\bf 76}, 3452 (1996)] derived a bound which reduces 
to the Holevo bound for complete measurements, but which is tighter for 
incomplete measurements. The most general quantum operations may be both 
incomplete and inefficient. Here we show that the bound derived by SWW can be 
further extended to obtain one which is yet again tighter for inefficient 
measurements. This allows us in addition to obtain a generalization of a bound 
derived by Hall, and to show that the average reduction in the von Neumann 
entropy during a quantum operation is concave in the initial state, for all 
quantum operations. This is a quantum version of the concavity of the mutual 
information. We also show that both this average entropy reduction and the mutual 
information for pure state ensembles, are Schur-concave for unitarily covariant 
measurements; that is, for these measurements, information gain increases with 
initial uncertainty.

\end{abstract}


\maketitle

\section{Introduction}

The celebrated Holevo bound, conjectured by Gordon~\cite{HBoundConj} and 
Levitin~\cite{Levitin}
and proved by Holevo in 1973~\cite{HBound} gives a bound on the information
which may be transmitted from A to B (strictly, the {\em mutual information},
$M$, between A and B) when A encodes information in a quantum
system using a set of states $\{\rho_i\}$, chosen with probabilities $\{P(i)\}$,
and B makes a subsequent measurement upon the system. The Holevo bound is
\begin{equation}
  M(I\! :\! J) \leq \chi \equiv S(\rho) - \sum_i P(i) S(\rho_i) ,
\end{equation}
where $\rho = \sum_i P(i) \rho_i$ (and which we will refer to as the {\em ensemble
state}). We write the mutual information as $M(I\! :\! J)$ to signify that it is the
mutual  information between the random variables $I$ and $J$, whose values $i$
and $j$ label  respectively the encoding used by A, and outcome of the
measurement made by B. More recent proofs of the Holevo bound may be found in
Refs.~\cite{YO93,FC94,SWW} The bound is achieved if and only if the encoding
states, $\rho_i$, commute with each other, and the  receiver, B, makes a von
Neumann measurement in the basis in which they are diagonal. (A von Neumann 
measurement is one that projects the system onto one of a complete set of 
mutually orthogonal states. In this case the set of states is chosen to be the 
basis in which the coding states are diagonal.) With this choice of coding states
and measurement the channel is classical, in that it can be implemented with a 
classical system. The Holevo bound takes into account that the sender
may only be able to send mixed states, and this mixing reduces the amount of
information that can be transmitted. However, if the receiver is not able to
perform measurements which always project the system to a pure state (so
called  {\em complete} measurements), then in general the information will be
further reduced. In 1996 Schumacher, Westmoreland  and Wootters showed that
when the receivers measurement is incomplete, it is possible to take this into
account and derive a more stringent bound on the information. If the receiver's
measurement is the POVM described by the operators $\{A_j\}$ (with $\sum_j
A_j^\dagger A_j = 1$), so that the measurement outcomes are labeled by the index
$j$, then the SWW bound is~\cite{SWW}
\begin{equation}
  M(I\! :\! J) \leq \chi - \sum_j P(j) \chi_j ,
  \label{SWW}
\end{equation}
where $P(j)$ is the probability of outcome $j$~\cite{note1}, and $\chi_j$ is the
Holevo quantity for the ensemble that the system remains in (from the point of
view of the receiver), given  outcome $j$. This bound can be at least partially 
understood by
noting that if the system still remains in some ensemble of possible states 
after the measurement, then future measurements can potentially extract further 
information about the encoding, and so the information obtained by the 
first measurement must therefore be less than the maximum possible by 
at least by this amount. What the SWW bound tells us is that the bound 
on the information is reduced 
not only by the amount of information which could be further extracted after outcome 
$j$ has been obtained, but by the Holevo bound on this information, $\chi_j$. 

If the initial state of the system is $\rho_i$, then after outcome $j$ the
final  state of the system is given by  $\tilde{\rho}_{j|i} = A_j \rho_i
A_j^\dagger/\mbox{Tr}[A_j^\dagger A_j \rho_i]$.  Thus the states which make up
the final ensemble that remains after outcome $j$, are $\{\tilde{\rho}_{j|i}\}$, 
and the probability of each state in the ensemble is $P(i|j) =
P(j|i)P(i)/P(j)$, with $P(j|i) = \mbox{Tr}[A_j^\dagger A_j \rho_i]$.
The Holevo quantity for ensemble $j$ is thus
\begin{equation}
  \chi_j = S(\tilde{\rho}_j) - \sum_i P(i|j) S(\tilde{\rho}_{j|i}),
\end{equation}
where $\tilde{\rho}_j = A_j \rho A_j^\dagger/\mbox{Tr}[A_j^\dagger A_j \rho]$. If
at least one of the measurement operators $A_j$ are higher than rank 1, then 
the measurement is incomplete. If the measurement is complete, then for each
$j$ all the final states $\tilde{\rho}_{j|i}$ are identical, $\chi_j$ is zero
and the SWW bound reduces to the Holevo bound. 

The most general kind of measurement can also be inefficient. A measurement is
described as inefficient if the observer does not have full information 
regarding which of the outcomes actually occurred. The name {\em inefficient} 
comes from that fact that the need to consider such measurements first arose 
in the study of inefficient photo-detectors.~\cite{inefficient} An inefficient
measurement  may be described by labeling the measurement operators with two
indices, so  that we have $A_{kj}$. The receiver has complete information
about one of the indices, $j$, but no information about the other,
$k$.~\cite{note2} As a result the final state for each $j$ (given the value of
$i$) is now
\begin{equation}
  \rho'_{j|i} = \sum_k P(k|j) \frac{A_{kj} \rho_i A_{kj}^\dagger}
                                    {\mbox{Tr}[A_{kj}^\dagger A_{kj} \rho_i]} .
\end{equation}
Since inefficiency represents a loss of information, we wish to ask whether it is
possible to take this into account and obtain a more stringent bound on the
mutual information. If we merely apply the SWW bound to the measurement
$A_{kj}$, then the bound involves the Holevo quantities of the ensembles that
remain when both the values of $k$ and $j$ are known (the final
ensembles that result from the efficient measurement). That is
\begin{equation}
  M(I\! :\! J) \leq \chi - \sum_{kj} P(k,j) \chi_{kj} .
\end{equation}
One therefore wishes to know whether it is possible to derive a bound which
instead involves the Holevo quantities of the ensembles that remain after the
inefficient measurement is made, that is, for the receiver who only has access to
$j$. 

In the first part of this paper we answer this question in the affirmative -
for an inefficient measurement where the known outcomes are labeled by
$j$, the bound given by Eq.(\ref{SWW}) {\em remains true}, where now the
$\chi_j$ are the Holevo quantities for the ensemble of states $\rho'_{j|i}$
which result from the inefficient measurement.

In the second part of the paper, we consider the average reduction in the von
Neumann entropy induced by a measurement:
\begin{equation}
 \langle \Delta S(\rho) \rangle \equiv S(\rho) - \sum_i P(j) S(\rho'_{j}).
\end{equation}

Here $\rho'_j$ is the state that results from outcome $j$, given that the 
initial state is $\rho$. Since the von Neumann entropy is a  measure of how much
we know about the state of the system, this is the  difference between what we
knew about the system state before we made the  measurement, and what we know
(on average) about the system state at the end of the measurement; it thus
measures how much we learn about the final state of the system. Equivalently, it
can be said to measure the degree of ``state-reduction'' which the measurement
induces. 

While it is the mutual information which is important for communication, the 
reduction in the von Neumann entropy is important for feedback control. 
Feedback control is the process of performing a sequence of measurements on 
a system, and applying unitary operations after each measurement in order 
control the evolution of the system. Such a procedure is useful for controlling 
systems which are driven by noise. If the ability to perform unitary 
operations is unlimited, then the von Neumann entropy provides a measure 
of the level of control which can be achieved: if the system 
has maximal entropy then the unitary operations have no effect on the system 
state whatsoever; conversely, if the state is pure then the system can be 
controlled precisely - that is, any pure state can be prepared. 
Thus the entropy measures the extent to which a pure state, or pure evolution 
can be obtained, and thus the level of predictability which can be achieved  
over the future behavior of the system~\cite{note3}.
The primary role of measurement in feedback control is therefore to reduce 
the entropy of the system. As such the average reduction in von Neumann entropy  
provides a ranking of the effectiveness of different measurements for feedback 
control, other things being equal. Further details regarding quantum 
feedback control and von Neumann entropy can be found in reference.~\cite{DJJ} 

The entropy reduction is also relevant to the transformation of pure-state 
entanglement, since the von Neumann entropy measures the entanglement of pure 
states. As a result this quantity gives the amount by which pure-state 
entanglement is broken by a local measurement.

We give two corollaries of the general information bound derived in the 
first part which involve $\langle
\Delta S(\rho) \rangle$. The first is a generalization of a bound derived by 
Hall~\cite{Hall,ib} to inefficient measurements. Hall's bound states that for 
efficient measurements the mutual information is bounded by 
$\langle\Delta S(\rho) \rangle$. We show that for inefficient measurements 
this becomes 
\begin{equation}
  M(I\! :\! J) \leq \langle\Delta S(\rho) \rangle - \sum_i P(i) \langle \Delta S(\rho_i)\rangle ,
\end{equation}
where $\langle \Delta S(\rho_i)\rangle$ is the average entropy reduction  
which would have resulted if the initial state had been $\rho_i$, and as above 
$\rho = \sum_i P(i)\rho_i$. 

The second is the fundamental property that, for all quantum operations, the average
reduction in von Neumann entropy is concave in the initial state $\rho$. That is
\begin{equation}
  \langle \Delta S(\rho) \rangle \geq \sum_i P(i) \langle \Delta S(\rho_i) \rangle .
\end{equation}

Finally, in the third part of this paper, we use the above result to show that for 
measurements which are uniform in their sensitivity across state-space (that is, 
measurements which are unitarily covariant), the amount which one learns about the 
final state always increases with the initial uncertainty, where this 
uncertainty is characterized by {\em majorization}. This is a quantum 
version of the much simpler classical result (which we also show) that the mutual 
information always increases with the initial uncertainty for classical measurements 
which are permutation symmetric. In addition we show that, for unitarily covariant 
measurements, the mutual information for pure-state ensembles also has this 
property. One can sum up these results by saying that the statement that 
information gain increases with initial uncertainty can fail to hold only if the
measurement is asymmetric in its sensitivity. 

\section{An information bound for general quantum operations}

We now show that the bound proved by SWW can be generalized to obtain a more
stringent bound for channels in which the receivers measurement is inefficient.
To show this it turns out that we can use the same method employed by SWW, but
with the addition of an extra quantum system which allows us to include the
inefficiency of the measurement. 

\newtheorem{theo}{Theorem}
\begin{theo}
For a quantum channel in which the encoding ensemble is $\varepsilon =
\{P(i),\rho_i\}$, and the measurement performed by the receiver is described  by
operators $A_{kj}$ ($\sum_{kj} A_{kj}^\dagger A_{kj} = 1$), where the measurement is
in general inefficient so that the receiver knows $j$ but not $k$, then the
mutual information, $M(I\! :\! J)$, is bounded such that  
\begin{equation}
   M(I\! :\! J) \leq  \chi  - \sum_j P(j) \chi_j ,
\end{equation}
where $P(j)$ is the overall probability for outcome $j$, 
$\chi = S(\rho) - \sum_i P(i) S(\rho_i)$ is the Holevo quantity for the
initial ensemble and
\begin{equation}
   \chi_j =  S(\sigma_{j})  - \sum P(i|j) S(\sigma_{j|i}) ,
\end{equation}
is the Holevo quantity for the ensemble, $\varepsilon_j$, that remains (from
the point of view of the receiver) once the measurement has been made, so that the
receiver has learned the outcome $j$, but not the value of $k$. Here the 
receiver's overall final state is 
\begin{equation}
   \sigma_{j} = \frac{\sum_{k} A_{kj} \rho A_{kj}^\dagger}{P(j)}  
              = \sum_{ik} P(i,k|j) \sigma_{kj|i} ,
\end{equation}
where $P(i,k|j)$ is the probability for both $i$ and outcome $k$ given $j$, and
$\sigma_{kj|i}$ is the final state that results given the initial state
$\rho_i$, and both outcomes $j$ and $k$. The remaining ensemble $\varepsilon_j
= \{P(i|j), \sigma_{j|i}\}$, where 
\begin{equation}
   \sigma_{j|i} = \sum_{k} P(k|j,i) \sigma_{kj|i}  
                = \frac{\sum_{k} A_{kj} \rho_i A_{kj}^\dagger}{P(j|i)} ,
\end{equation}
and where $P(k|j,i)$ is the probability for outcome $k$ given $j$ and the 
initial state $\rho_i$.  
\end{theo}

\begin{proof}
We begin by collecting various key facts. The first is that any efficient
measurement on a system $Q$, described by $N = N_1N_2$ operators, $A_{kj}$, 
($j=1,\ldots , N_1$ and $k=1,\ldots , N_2$) can be obtained by bringing up an
auxiliary system $A$ of dimension $N$, performing a unitary operation
involving $Q$ and $A$, and then making a von Neumann measurement on
$A$.~\cite{Krauss,Schumacher} If the initial state of $Q$ is  $\rho^{(Q)}$,
then the final joint state of $A$ and $Q$ after the von Neumann measurement is 
\begin{equation}
   \sigma^{(AQ)} = |kj\rangle \langle kj|^{(A)} \otimes \frac{A_{kj} \rho^{(Q)} A_{kj}^\dagger}
                                                              {P(k,j)} .
\end{equation} 
where $|kj\rangle$ is the state of $A$ selected by the von Neumann measurement. 
The second fact is that the state which results from discarding all information 
about the measurement outcomes $k$ and $j$ can be obtained by performing a unitary operation between $A$ and another system $E$ which perfectly correlates the states $|kj\rangle$ of $A$ with orthogonal states of $E$, and then tracing out $E$.
The final
key fact we require is a result proven by SWW~\cite{SWW}, which is that the Holevo
$\chi$ quantity  is non-increasing under partial trace. That is, if we have two
quantum systems  $A$ and $B$, and an ensemble of states $\rho_i^{(AB)}$ with
associated probabilities $P_i$, then 
\begin{eqnarray}
   \chi^{(A)} & =    & S(\rho^{(A)}) - \sum_i S(\rho_i^{(A)})  \nonumber \\
              & \leq & S(\rho^{(AB)}) - \sum_i S(\rho_i^{(AB)}) = \chi^{(AB)} ,
\end{eqnarray}
where $\rho_i^{(A)} = \mbox{Tr}_B[\rho_i^{(AB)}]$. To prove this result  
SWW use strong subadditivity.~\cite{subadd} 

We now encode information in system $Q$ using the ensemble $\varepsilon$, and 
consider the joint system which consists of the three systems $Q$, $A$, $E$ and
a forth system $M$, with dimension $N_1$. We now start with $A$, $E$ and $M$
in pure states,  so that the Holevo quantity for the joint system is
$\chi^{(QAEM)} = \chi^{(Q)}$.  We then perform the required unitary operation
between $Q$ and $A$, and a unitary  operation between $A$ and $E$ which
perfectly correlates the states $|kj\rangle^{(A)}$ of $A$ with orthogonal states 
of $E$. Unitary operations do not change the Holevo quantity. Then we 
trace over $E$, so that we are left with the state
\begin{equation}
   |\psi\rangle \langle\psi|^{(M)} \otimes \sum_{jk} P(k,j) |k,j\rangle \langle k,j|^{(A)} \otimes \frac{A_{kj} \rho^{(Q)} A_{kj}^\dagger}{P(k,j)} .
\end{equation}
After the two unitaries and the partial trace over $E$, the Holevo quantity for the remaining systems, which we will denote by $\chi'^{(QAM)}$, satisfies $\chi'^{(QAM)} \leq \chi^{(QAEM)} = \chi^{(Q)}$. We now 
perform one more unitary operation, this time between $M$ and $A$, so that 
we correlate the states of $M$, which we denote by $|j\rangle\langle j|^{(M)}$ 
with the second index of the states of $A$, giving
\begin{equation}
  \sum_j |j\rangle\langle j|^{(M)} \otimes \sum_{k} P(k,j) |k,j\rangle \langle k,j|^{(A)} \otimes \sigma_{kj}^{(Q)}
\end{equation}
where $\sigma_{kj}^{(Q)} = A_{kj} \rho^{(Q)} A_{kj}^\dagger/P(k,j)$ is the
final state  resulting from knowing both outcomes $k$ and $j$, with no
knowledge of the initial  choice of $i$. Finally we trace out $A$, leaving us
with the state 
\begin{equation}
  \sigma^{(QM)} = \sum_j |j\rangle\langle j|^{(M)} \otimes \sum_{k} P(k,j) \sigma_{kj}^{(Q)}
\end{equation}

After this final unitary, and the partial trace over $A$, the Holevo quantity for the remaining systems $Q$ and $M$, which we will denote by $\chi''^{(QM)}$, satisfies $\chi''^{(QM)} \leq \chi'^{(QAM)} \leq \chi^{(Q)}$. We have gone 
through the above process
using the initial state $\rho$, but we could just as easily have
started with any of the initial states, $\rho_i$, in the ensemble, and we will
denote the final states which we obtain using the initial state $\rho_i$ as
$\sigma_i^{(QM)}$. Calculating $\chi''^{(QM)}$ we have 

\begin{eqnarray}
  \chi''^{(QM)} & = & S(\sigma^{(QM)}) - \sum_i P(i) S(\sigma_i^{(QM)}) \nonumber \\ 
              & = & H[J] - \sum_i P(i) H[J|i] \nonumber \\
              &   & + \sum_j P(j) \left[ S(\sigma_j) - \sum_i P(i|j) \sigma_{j|i} \right] \\
	      & = & M(J:I) + \sum_j P(j) \chi_j^{(Q)} \leq \chi^{(Q)}.
\end{eqnarray}
Rearranging this expression gives the desired result.
\end{proof}

\section{Properties of entropy reduction}

We now rewrite the above information bound using the fact that $P(i|j)P(j) =
P(j|i)P(i)$. The result is 
\begin{equation}
  M(I\! :\! J) \leq \langle \Delta S(\rho)\rangle - \sum_i P(i) \langle \Delta S(\rho_i)\rangle
\label{genHall}
\end{equation}
where $\rho = \sum_i P_i \rho_i$. Ozawa has shown that
for efficient measurements $\langle \Delta S(\rho) \rangle$ is always positive\cite{Ozawa} 
(for more recent proofs of this result see\cite{Nielsen,FJ}).
For efficient measurements Eq.(\ref{genHall}) is therefore in general stronger than,
and gives immediately, 
Hall's bound~\cite{Hall,ib}, which states that the mutual information is
bounded by the reduction in the von Neumann entropy. The inequality in Eq.(\ref{genHall}) is
then a generalization of Hall's bound to inefficient measurements. Since the
mutual information is always positive, but for inefficient measurements the
reduction in the von Neumann entropy can be negative (that is the entropy of the 
quantum state can {\em increase} as a result of the measurement), the relation
\begin{equation}
  M(I\! :\! J) \leq \langle \Delta S(\rho)\rangle 
\end{equation}
is not necessarily satisfied for such measurements. However, Eq.(\ref{genHall})
tells  us that if the entropy of the intial state, $\rho$, does increase, the 
average increase in the entropy for each of the coding states $\rho_i$ is always 
{\em more} that this by at least the mutual information.

The second result that we obtain from Eq.(\ref{genHall}) is that, because the 
mutual information is nonnegative, we have 
\begin{equation}
  \langle \Delta S(\rho)\rangle \geq \sum_i P(i) \langle \Delta S(\rho_i)\rangle .
\end{equation}
That is, the reduction in the von Neumann entropy is concave in the initial 
state. This parallels the fact that the mutual information is also concave 
in the initial state. 

The fact that this is true for inefficient measurements, means that once we
have  made an efficient measurement, no matter what information we throw
away regarding the final outcomes (i.e. which outcomes we average over),
$\langle \Delta S(\rho)\rangle$ is always greater than 
the average of the entropy reductions
which would have been obtained through measurement in each of the coding
states, when we throw away the same information regarding the measurement
results.
 


\section{Information gathering and state-space symmetry}
In this section we show that measurements whose ability to 
extract information is uniform over the available state-space 
(that is, does not vary from point to point in the state-space) 
always extract more information (strictly, never
extract less information) the less that is known before the measurement is made.
Thus, in this sense, one may regard ``the more you know, the less you get'' as a 
fundamental property of measurement. We will show that this is true both for 
the information obtained regarding the final state (being $\langle \Delta
S(\rho)\rangle$), and the mutual information for a measurement on an ensemble
of pure states. We will consider here efficient measurements only; no doubt 
inefficient measurements will also have this property, but only if 
the information which is thrown away is also uniform with respect to the 
state-space, and we do not wish to burden the treatment with this additional 
complication.

To proceed we must make precise the notion that the sensitivity of a measurement is uniform over state-space. This is captured by stating that such a
measurement should be invariant under reversible transformations of the
state-space. For classical measurements (which are simply quantum measurements
in which all operators and density matrices commute~\cite{pool}) this means 
that the set of measurement operators is invariant under all permutations of 
the classical states: we will refer to these as {\em completely symmetric} measurements. Note that in this classical case, this is equivalent to saying that the measurement distinguishes all states from all other states equally well.
The quantum generalization of this is invariance under all unitary
transformations. Such measurements are referred to as being unitarily covariant.~\cite{Barnum,Ucov}

We must also quantify what we mean by the observer's lack of knowledge,  or
uncertainty, before the measurement is made. This is captured by the simple and
elegant concept of {\em majorization}.~\cite{MO,Bhatia} If two sets of
probabilities $p \equiv \{P_i\}$ and $q \equiv \{Q_i\}$ satisfy the set of
relations 
\begin{equation}
   \sum_{i=1}^{k} P_i \geq \sum_{i=1}^{k} Q_i \;\; , \;\;\; \forall k,
\label{majdef}
\end{equation}
where it is understood that the elements of both sets have been placed in
decreasing order (e.g., $P_i > P_{i+1}, \forall i$), then $p$ is said to
majorize $q$, and this is written $q \prec p$. While at first Eq.(\ref{majdef})
looks a little complicated, a few moments consideration reveals that it
captures precisely what one means by uncertainty - if $p$ majorizes $q$, then
$p$ is more sharply peaked than $q$, and consequently describes a state of
knowledge containing less uncertainty. What is more, majorization implies an
ordering with Shannon entropy $H[\cdot]$. That is, if $p$ majorizes $q$, then
$H[p] \leq H[q]$.~\cite{MO,Bhatia} 

In a sense, majorization is a more basic notion of uncertainty than entropy in
that it captures that concept alone -- the Shannon entropy on the other hand
characterizes the more specific notion of information. 
To characterize the uncertainty of a density matrix, we can apply majorization
to the vector consisting of its eigenvalues. If $\rho$ and $\sigma$ are density
matrices, then we will write $\sigma \prec \rho$ if $\rho$'s eigenvalues
majorize $\sigma$'s. Various applications have been found for majorization in
quantum information theory.~\cite{NielsenLett,JPV,NK,Nielsen,Chefles,FJ}

We thus desire to show that for measurements with the specified symmetry, 
$\langle \Delta S(\sigma)\rangle \geq \langle \Delta S(\rho)\rangle $ whenever
$\sigma \prec \rho$ (and similarly for the mutual information). Functions with
this property (of which the von Neumann entropy, $S(\rho)$, is one example) are
referred to as being {\em Schur-concave}. To show that a function is
Schur-concave, it is sufficient  to show that it is concave, and symmetric in
its arguments~\cite{MO,Bhatia}, which in our case are the eigenvalues of the
density matrix $\rho$ (if our functions did not depend only on the eigenvalues
of $\rho$, then they could not be Schur-concave, since the majorization
condition only involves these eigenvalues).  

The desired result for classical completely symmetric measurements is now
immediate. In the classical case the mutual information is the unique measure
of information gain, and $ M(I\! :\! J) = \langle \Delta S(\rho)\rangle$. The mutual
information is concave in the initial classical probability vector ${\bf P} =
(P_1,\ldots,P_n)$ (being the vector of the eigenvalues of $\rho$ in our quantum
formalism), as is indeed implied by the concavity of $\langle \Delta
S(\rho)\rangle$. Since all operators commute with the density matrix, $\langle
\Delta S(\rho)\rangle$ is only a function of the $\{P_i\}$. From the form of
$\langle \Delta S(\rho)\rangle$ we see that a permutation of the elements of
${\bf P}$ is equivalent to a permutation applied to the measurement operators,
and since these are invariant under such an operation, $\langle \Delta
S(\rho)\rangle$, and thus $M(I\! :\! J)$, is a symmetric function of its arguments.
Thus $M(I\! :\! J)$ is Schur-concave.

The Schur-concavity of $\langle \Delta S(\rho)\rangle$ for unitarily covariant 
(UC) quantum measurements is just as immediate. Because of the unitary covariance
of the measurement, we see from the form of $\langle \Delta S(\rho)\rangle$
that it is invariant under a unitary transformation of $\rho$. As a result, it
only depends upon the eigenvalues of $\rho$. Since the permutations are a
subgroup of the unitaries, it is also a symmetric function of its arguments (the
eigenvalues), and thus Schur-concave.

We wish finally to show that the mutual information is also Schur-concave in 
$\rho$ for unitarily covariant measurements on ensembles of pure states. This 
requires a little more work. First we need to show that once we have fixed a 
set of encoding states, the mutual information 
is concave in the vector of the ensemble probabilities $P(i)$. This is straightforward 
if we first note that the
mutual information, because it is, in fact, symmetric between $i$ and $j$, can
be written in the reverse form
\begin{eqnarray}
  M(I\! :\! J) & = & H[P(j)] - \sum_j P(i) H[P(j|i)] ,
  \label{revMI}
\end{eqnarray}
Since, for a fixed measurement, the mutual information is a function of the 
ensemble probabilities we will write it as $M(\{P(i)\})$.
Denoting the pure states in the encoding ensemble as $\rho_i =
|\psi_i\rangle\langle\psi_i|$, and choosing the ensemble state $\rho = \sum_k P_k \sigma_k$, where the
$\sigma_k$ are built from the encoding states so that $\sigma_k = \sum_i
P_{i|k} |\psi_i\rangle\langle\psi_i|$, then 
\begin{eqnarray}
  & & M(\{P(i)\}) \nonumber \\
                  & = & H[\sum_k P(k) P(j|k)] - \sum_i \sum_k P(i|k) P(k) H[P(j|i)] \nonumber \\
                & \ge & \sum_k P(k) H[P(j|k)] - \sum_k P(k) \sum_i P(i|k) H[P(j|i)] \nonumber \\
                  & = & \sum_k P(k) M(\{P(i|k)\}) ,
\label{cconc}
\end{eqnarray}
being the desired concavity relation. The inequality in the third line is
merely a result of the concavity of the Shannon entropy. Note that while we 
have written the measurement's outcomes explicitly as being discrete in the 
about derivation, the result also follows if they are a continuum (as in 
the case of UC measurements) by replacing the relevant sums with integrals. 

Now we need to note some further points about UC measurements: A UC measurement
may be generated by taking
all unitary transformations of any single operator $A$,
and dividing them by a common normalization factor. The resulting measurement operators are thus $A_U \propto UAU^\dagger$, where  $U$ ranges over all unitaries. The
normalization for the $A_U$ comes from $\int U A^\dagger A U^\dagger d\mu(U) = \mbox{Tr}[A^\dagger A] I$ where $d\mu(U)$ is the (unitarily invariant) Haar measure~\cite{Ucov,Jones} over unitaries. 

It is not hard to show that all UC measurements can be obtained by {\em
mixing} different UC measurements, each generated by a different operator.
(Mixing a set of measurements means assigning to each a probability, and then
making one measurement from the set at random based on these
probabilities~\cite{mixing}). 


Next, we need to show that for all UC measurements the mutual information depends only on the eigenvalues of the ensemble density matrix, and we state this as the following lemma.

\newtheorem{lem1}{Lemma}
\begin{lem1} The mutual information for a UC measurement on a
pure-state ensemble, $\varepsilon = \{P(i),|\psi_i\rangle\}$ depends on the ensemble only through the eigenvalues of the
density matrix $\rho=\sum_{i}P(i)|\psi_{i}\rangle\langle \psi_{i}|$.
\end{lem1}

\begin{proof} 
We first show this for UC measurements generated from a single operator. Writing the mutual information in the reverse form one has
\begin{eqnarray}
	 M(I\! :\! J)) & = & H[P(U)] - \sum_i P_i H[P(U|i)] ,
\label{forIc}
\end{eqnarray}
where $U$ is the continuum index for the measurement operators (and thus the
measurement outcomes)  which are $A_U = UAU^\dagger$ for some appropriately 
normalized $A$. Naturally all
this means is that $P(U|i)$ is a function of $U$, where $U$ ranges over all
unitaries. Since the measurement is unitarily covariant, $H[P(U|i)]$ is the
same for all initial states $|\psi_{i}\rangle$, and therefore the second term
is the same for all initial ensembles.  Thus $M$ depends only on the
first term $H[P(U)] = H[\mbox{Tr}[UA^\dagger AU^\dagger\rho]]$, which depends only on
$\rho$, and is invariant under all unitary transformations of $\rho$. Thus
$M$ depends only on the eigenvalues of $\rho$. Since the mutual information 
for a mixture of measurements is merely a function of the respective mutual informations for each measurement (in particular it is a linear combination of them), the result holds for all UC measurements.
\end{proof}

Since $M$ depends only on $\rho$, in establishing the Schur concavity of
$M$ with respect to $\rho$, we need only consider one ensemble for each
$\rho$. We therefore choose the eigen-ensemble $\{\lambda_i,|\phi_i\rangle\}$,
where $\lambda_i$ and $|\phi_i\rangle$ are the eigenvalues and eigenvectors of
$\rho$ respectively. We know that the mutual information is concave in the
vector of initial ensemble probabilities, and for the ensemble we have chosen,
the initial probabilities are the eigenvalues of $\rho$. As a result the mutual
information is concave in the eigenvalues of $\rho$. Since $M$ is invariant under 
unitary transformations, and since unitary transformations include 
permutations as a subgroup, it is also a symmetric function of the eigenvalues. 
Thus $M$ is Schur-concave.

\section{conclusion}
In using a quantum channel, if there are limitations on the completeness (or alternatively the {\em strength}, in the terminology of~\cite{FJ}) or efficiency of the measurements that the receiver can perform, then it is possible to give a bound on the mutual information which is stronger than the Holevo bound. Further, this bound has a very simple form in terms of the Holevo $\chi$ quantity, and the $\chi$ quantities of the ensembles, one of which remains after the measurement is made. 

This bound also allows us to obtain a relationship between the mutual information and the average von Neumann entropy reduction induced by a measurement, and encompasses the fact that this von Neumann entropy reduction is concave in the initial state.  

From the concavity of the mutual information and the von Neumann entropy reduction, it follows that these quantities are Schur-concave (the former naturally for pure-state ensembles) for completely symmetric classical measurements, and for unitarily covariant quantum measurements. Thus the possibility that either of
these kinds of information gain {\em decreases} with increasing initial uncertainty is associated with the asymmetry of the measurement in question.

\section*{Acknowledgments}
The author would like to thank Gerard Jungman, Howard Barnum,
Howard Wiseman, Terry Rudolph and Michael Hall for helpful discussions. 
The author is also grateful both to
Vlatko Vedral for hospitality during a visit to Imperial College, and Lucien
Hardy for hospitality during a visit to the Perimeter Institute where some of
the initial stages of this work was carried out. This work was supported by
the Australian Research Council and the State of Queensland.
Note added: After submitting this manuscript, which was first posted as eprint 
quant-ph/0412006, it was brought to my attention that the work presented 
here overlaps with concurrent work by Barchielli and Lupeiri (quant-ph/0409019 and 
quant-ph/0412116).


\begin{thebibliography}{99}
\bibitem{note1}
   While it is an abuse of notation to denote the ensemble probabilities by 
   $P(i)$, and the (in general unrelated) outcome probabilities by $P(j)$, 
   we use it systematically throughout, since we feel it keeps the notation 
   simpler, and thus ultimately clearer.
\bibitem{note2}
   If the observer has only partial information about the outcome of a
   measurement, then if we label the outcomes by $n$ (with associated 
   measurement operators $B_n$), the most general situation is one in which the
   observer knows instead the value of a second variable $m$, where $m$ is
   related to $n$ by an arbitrary conditional probability $P(m|n)$. This
   general case is encompassed by the two-index formulation we use in the text.
   To see this, one sets $k=n$, $j=m$ and chooses $A_{nm} (\equiv A_{kj}) =
   \alpha_{nm} B_{n}$. Then by giving the observer complete knowledge of $j$,
   and no knowledge of $k$, we reproduce precisely the general case described
   above by choosing $\alpha_{nm}$ so that $|\alpha_{nm}|^2 = P(m|n)$.
\bibitem{note3}
   The von Neumann entropy is not the only quantity which can be used to measure 
   the achieved level of control. 
   The von Neumann entropy specifically gives the minimum possible entropy of 
   the results of a measurement on the system. It therefore measures 
   the maximum information (strictly, the minimum information deficit) which 
   the user who is performing the control has about of the future behavior 
   of the system under measurement. An example of another measure of control 
   is the maximum eigenvalue of the density matrix. Under the assumption that 
   all unitary operations are available to the controller, this measures the 
   probability that the controlled system will be found in the desired state.
\end{thebibliography}
\end{document}